# Type-Based Detection of
# XML Query-Update Independence

Nicole Bidoit-Tollu
Universite Paris Sud &
INRIA Saclay
bidoit@lri.fr

Dario Colazzo
Universite Paris Sud &
INRIA Saclay
colazzo@lri.fr

Federico Ulliana
Universite Paris Sud &
INRIA Saclay
fulliana@lri.fr

## ABSTRACT

This paper presents a novel static analysis technique to detect XML query-update independence, in the presence of a schema. Rather than types, our system infers *chains* of types. Each chain represents a path that can be traversed on a valid document during query/update evaluation. The resulting independence analysis is precise, although it raises a challenging issue: recursive schemas may lead to inference of infinitely many chains.

A sound and complete approximation technique ensuring a finite analysis in any case is presented, together with an efficient implementation performing the chain-based analysis in polynomial space and time.

## 1. INTRODUCTION

A query and an update are independent when the query result is not affected by update execution, on any possible input database. Detecting query-update independence is of crucial importance in many contexts: i) to minimize view re-materialization; ii) to ensure isolation, when queries and updates are executed concurrently; iii) as outlined in [6], to enforce access control policies, when the query is used to define the part of the database that must not be changed by a user update.

In all these contexts, benefits are amplified when query-update independence can be checked statically. In order to be useful, every static analysis technique must be sound: if query-update independence is statically detected, then independence does hold. The inverse implication (completeness) cannot be ensured in the general case, since static independence detection is undecidable (see [6]). This means that if a static analyzer is used, for instance, in a view maintenance system, sometimes views are re-materialized after updates even if not needed, because the analysis has not been smart enough to statically detect a view-update independence. Useless view re-materialization frequently occurs if a static analyzer with low precision is adopted. This can lead to great waste of time, since view materialization cost can be proportional to the database size.

High precision of static independence analysis can be ensured by taking into account schema information. In many contexts, schemas are defined by users, mainly by means of the DTD or XML

Schema languages, while in other contexts quite precise schemas, in the form of a DTD, can be automatically inferred, by using accurate and efficient existing techniques like the one proposed by Bex et al. in [8].

### State of the Art

Schema-based detection of XML query-update independence has been recently investigated. The state of the art technique has been presented by Benedikt and Cheney in [6]. This technique infers from the schema the set of node types traversed by the query, and the set of node types impacted by the update. The query and the update are then deemed as independent if the two sets do not overlap. This technique is effective since the static analysis i) is able to manage a wide class of XQuery queries and updates, ii) can be performed in a negligible time, and iii) as a consequence, even on small documents, can avoid expensive query re-computation when independence wrt an update is detected. However, the technique has some weaknesses. As illustrated in [6], in some cases, independence is not detected, due to some over-approximation made by the type inference rules.

For example, this technique cannot detect independence between the query $q_1 = //a//c$ and the update $u_1 = $ delete $//b//c$, when the schema enforces that $c$ descendants of $b$ nodes are never descendants of $a$ nodes. This is because the type inference technique of [6] infers the type $c$ both for the query path and the update path, without considering contextual information about the inferred types. Since the query and update types overlap, independence is wrongly excluded. Indeed, the technique is not precise enough when ancestor or descendant axes are used in queries and updates.

The way XPath axes are typed is not the only source of low precision of this technique. Consider documents typed by the well known bibliographic DTD used in [1], the query $q_2 = //title$ and the update $u_2 = $ for x in $//book$ return insert $<author/>$ into x. The technique of [6] infers *bib*, *book* and *title* as types traced by $q_2$, and *book* as type impacted by $u_2$. According to this technique, the two expressions share the type *book*, hence independence is not detected, while it holds.

In none of the above examples, independence can be detected by techniques ignoring schema information like the path-based approach proposed by Ghelli et al.[1] [15] and the recent destabilizers-based approach proposed by Benedikt and Cheney [5]. Following these approaches, for the example $q_1$-$u_1$, the paths $//a//c$ and $//b//c$ are deemed as overlapping since, for instance, documents matching the path $/a/b/c$ match both paths and similarly, for example $q_2$-$u_2$ and the paths $//title$ and $//book$.

---

[1] This technique deals with update-commutativity detection for a language with side effects and can be directly extended to query-update independence detection without side effects.

## Contributions

This paper proposes a novel schema-based approach for detecting XML query-update independence. Differently from [6, 10, 11], our system infers sequences of labels (hereafter called *chains*). Intuitively, for each node that can be selected by a query/update path in a schema instance, the system infers a chain recording i) all labels that are encountered from the root to the node, ii) in the order of traversal. This information is at the basis of a precise static independence analysis. For instance, for $q_1 = //a//c$ and $u_1 = //b//c$ over the schema $\{ doc \leftarrow (a|b)*,\ a \leftarrow c,\ b \leftarrow c \}$, the chains $doc.a.c$ and $doc.b.c$ are inferred for the query and the update, respectively. Disjointness of these two chains allows us to statically derive the independence for $q_1$-$u_1$. For the DTD of the XQuery Use Cases [1] before discussed, the chains $bib.book.title$ and $bib.book.author$, respectively inferred for $q_2$ and $u_2$, diverge after the *book* symbol; this allows us to conclude independence for $q_2$-$u_2$. These two examples highlight that chain inference provides a more precise independence analysis than that of [6, 15, 5].

The main contribution of the paper is a precise algorithm to detect independence for a query-update pair q-u knowing that documents are valid wrt a DTD d. It strongly relies on the following developments.

- Chain-based independence for q-u, a static notion, is the foundation of our algorithm: starting from the set $C_d$ of all possible chains associated with the DTD d, our inference system extracts subsets of chains $C_q$ and $C_u$ which soundly approximate the navigation through valid documents made by the evaluation of the query q and the update u, respectively. Note that our inference system is cautiously specified for dealing with all XPath axes. Chain-based independence is the result of the absence of overlapping pair of chains in $C_q$ and $C_u$. Chain-based independence is proved to be sound wrt the semantics notion of query-update independence (Section 4).

- A major step of our work concerns recursive schemas, for which chain-based independence analysis may cripplingly involve to deal with an infinite number of chains. Our technique enabling the restriction of the analysis to finite subsets of $C_q$ and $C_u$ is a key contribution, and the core of our algorithm is the resulting finite analysis (Section 5). It is proved to be equivalent to the *infinite* analysis [2].

- The algorithm has been carefully implemented, and extensive tests have been performed to validate our claim of precision and efficiency (Section 6). Indeed, using a DAG-based representation of inferred chains allows the finite analysis to run in polynomial space and time. Concerning precision, our results show that our technique outperforms [6] to a large extent. Test results also show that high savings of time can be ensured by avoiding re-evaluation of queries deemed as independent of an update, even on relatively small documents.

A nice property of our technique (Section 7) is that it can be easily extended in order to cope with Extended DTDs [14], and thus XML Schema. Discussions about related and future work are provided in Sections 8 and 9.

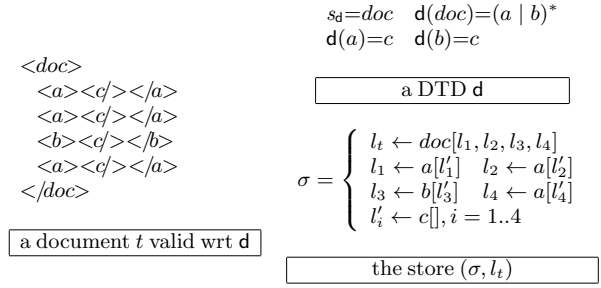

$<doc>$
$<a><c></c></a>$
$<a><c></c></a>$
$<b><c></c></b>$
$<a><c></c></a>$
$</doc>$

a document $t$ valid wrt d

$s_d = doc \quad d(doc) = (a \mid b)^*$
$d(a) = c \quad d(b) = c$

a DTD d

$\sigma = \begin{cases} l_t \leftarrow doc[l_1, l_2, l_3, l_4] \\ l_1 \leftarrow a[l_1'] \quad l_2 \leftarrow a[l_2'] \\ l_3 \leftarrow b[l_3'] \quad l_4 \leftarrow a[l_4'] \\ l_i' \leftarrow c[], i = 1..4 \end{cases}$

the store $(\sigma, l_t)$

**Figure 1: Document and store**

## 2. PRELIMINARIES

*Data model.* We represent an instance of the XML data model as a store $\sigma$, which is an environment associating each node location (or identifier) $l$ with either an element node $a[L]$ or a text node s. In $a[L]$, $a$ is the element tag, while $L = (l_1, \ldots, l_n)$ is the ordered list of children locations in $\sigma$. A tree is a pair $t = (\sigma, l_t)$, where $l_t$ is its root location. $dom(\sigma)$ denotes the set of locations of $\sigma$, while $\sigma@l$ denotes the subtree of $\sigma$ rooted at $l$ whose domain is limited to locations connected to $l$. See Figure 1 for a small document together with its store.

*DTDs.* A DTD is a 3-tuple $(\Sigma, s_d, d)$ where: $\Sigma$ is a finite alphabet for element tags, denoted by $a, b, c$; $s_d \in \Sigma$ is the start symbol; d is a function from $\Sigma$ to regular expressions over $\Sigma \cup \{S\}$, where S denotes the string type. For simplicity, next we often use only the d component to specify a DTD.

A tree $t = (\sigma, l_t)$ is valid wrt d, denoted $t \in d$, iff there exists a mapping $\nu : dom(t) \mapsto \Sigma \cup \{S\}$ such that: $\nu(l_t) = s_d$; $\nu(l) = S$ implies that $\sigma(l)$ is a text node; $\nu(l) = a$ implies that $\sigma(l) = a[L]$ and the word $\nu(L)$ is generated by the regular expression $d(a)$.

**DEFINITION 2.1** (REACHABILITY AND CHAINS). *Let* d *be a DTD,* $\alpha \Rightarrow_d \beta$ *holds iff* $\alpha, \beta \in \Sigma \cup \{S\}$ *and* $\beta$ *occurs in the regular expression* $d(\alpha)$. *A chain* c *over* d *is a sequence of labels* $\alpha_1.\alpha_2 \ldots \alpha_n$ *such that* $\alpha_i \Rightarrow_d \alpha_{(i+1)}$ *for* $i = 1 \ldots n-1$.
*The set of chains associated with the DTD* d *is denoted* $C_d$.

For the DTD of Figure 1, the set $C_d$ includes the chains $doc.a$, $a.c$, $doc.a.c$, $doc.b$, $b.c$, and $doc.b.c$, because we have $doc \Rightarrow_d a$, $a \Rightarrow_d c$, $doc \Rightarrow_d b$ and $b \Rightarrow_d c$.

Given two chains $c_1$ and $c_2$, the concatenation of $c_1$ and $c_2$ is denoted $c_1.c_2$; we write $c_1 \preceq c_2$ to indicate that $c_1$ is a prefix of $c_2$, that is $c_2 = c_1.c'$ for some chain $c'$.

Observe that chains in $C_d$ are of finite length and may start with any DTD symbol. The set $C_d$ is infinite only if d is a vertical-recursive schema.

**DEFINITION 2.2** (NODE TYPE AND CHAIN). *Given* $\sigma$ *and* $l \in dom(\sigma)$, *we define* $\mathsf{typ}(l) = a$ *if* $\sigma(l) = a[L]$, *otherwise* $\mathsf{typ}(l) = S$. *The chain associated to the node* $l$ *is defined by* $c_l^\sigma = \mathsf{typ}(l)$ *if* $l$ *has no parent, otherwise* $c_l^\sigma = c_{parent(l)}^\sigma.\mathsf{typ}(l)$.

Consider the DTD and store of Figure 1, we have:
$\mathsf{typ}(l_1) = \mathsf{typ}(l_2) = \mathsf{typ}(l_4) = doc.a$ and
$\mathsf{typ}(l_1') = \mathsf{typ}(l_2') = \mathsf{typ}(l_4') = doc.a.c$.

**PROPOSITION 2.3.** *Given a tree* $t = (\sigma, l_t) \in d$, *for each location* $l \in dom(\sigma)$, *we have* $c_l^\sigma \in C_d$.

## Queries, Updates and Independence.

We assume that the reader is familiar with the XQuery and XQuery Update Facility languages. In this paper we consider the two large fragments considered in related approaches [6, 5], respectively defined by the following grammars.

```
q ::=  ()  |  q.q  |  <a>q</a>  |  s  |  x/step
       | for x in q return q  |  let x := q return q
       | if q then q else q

step ::=  axis :: φ          φ ::= a  |  text()  |  node()

axis ::=  self  |  child  |  descendant
          | descendant−or−self  |  parent
          | ancestor  |  ancestor−or−self
          | preceding−sibling  |  following−sibling
```

The empty-sequence and sequence queries are denoted by () and q.q respectively. The query s denotes a constant string value. The symbol $\phi$ is used for XPath node tests; $a$ stands for a tag symbol. XPath expressions $x/step_1 / \ldots /step_n$, although used in examples, are not directly supported by the grammar; they can be encoded in the standard way, by means of iteration and the allowed single step expression; axes that are not included can be easily encoded too.[3] The rest of the grammar is self explicative.

In examples, $/\phi$ and $//\phi$ are respectively used as shortcuts for /child::$\phi$ and /descendant−or−self::node()/child::$\phi$.

Also, to simplify the formal treatment, we assume that element construction $<a>q</a>$ is not used in the left-hand side expression of a for/let-expression. This restriction is met by a very large class of queries used in practice, while queries like let x := <a>q'</a> return <b>x</b> can be rewritten by simple variable substitution.

The subset of XQuery Update Facility we consider is defined as follows. All update operations (namely, insert, delete, rename and replace) are included.

```
u    ::=  ()  |  u,u  |  for x in q return u
          | let x := q return u
          | if q then u1 else u2
          | delete q0  |  rename q0 as a
          | insert q pos q0  |  replace q0 with q

pos  ::=  before  |  after  |  into (as first | as last)?
```

Like for queries, updates can be composed sequentially or by means of let/for statements, where only the return part can contain update operations. In other update expressions, $q_0$ is the *target* expression producing the (*target*) node in the input document, that is where the update has to be done. In insert and replace updates, q is the *source* expression producing elements for the insertion or replacement. Deletion delete $q_0$ and renaming rename $q_0$ as $a$ are self-explicative. According to the W3C semantics [19] the target expression $q_0$ is required to output a single node otherwise a run time error occurs.

Query and update semantics are specified in [12, 19], while a succinct and elegant formalization can be found [4], from which we borrow some notions that are needed for our own presentation. Query semantics is denoted by

$$\sigma, \gamma \models q \Rightarrow \sigma_q, L_q$$

meaning that the execution of the query q over $\sigma$ outputs a sequence of locations $L_q$, roots of the answer trees for q, and a new store $\sigma_q$,

---

[3] e.g., /following::$a$ becomes /ancestor−or−self::node()/ following−sibling::node()/descendant−or−self::$a$.

---

including $\sigma$ plus new elements built by q; the environment $\gamma$ binds each free variable of q to a sequence of locations in $\sigma$.

According to the W3C specification, update evaluation is split into three phases: i) creation of an *update pending list* (UPL) of simple update commands, ii) execution of sanity check on this list, and iii) application of the UPL on the input store so as to obtain the updated data. Update commands $\iota$ in a UPL $w$ are of the form:

$$\iota ::= \texttt{ins}(L, pos, l)  |  \texttt{del}(l)  |  \texttt{repl}(l, L)  |  \texttt{ren}(l, a)$$

where $l$ is the *target location* and $L$ the sequence of roots of source elements to be inserted. The creation of the UPL (phase i) is denoted by:

$$\sigma, \gamma \models u \Rightarrow \sigma_w, w$$

As usual, $\gamma$ binds $u$ free variables to locations in $\sigma$ and the store $\sigma_w$ contains newly created locations potentially used in the UPL $w$. Applying the UPL $w$ to the input store $\sigma$ (phase iii) produces the updated store. This is denoted by:

$$\sigma_w \vdash w \rightsquigarrow \sigma_u$$

The composition of phases i) and iii) is denoted by:

$$\sigma, \gamma \models u : \sigma_u$$

Above, $dom(\sigma) \subseteq dom(\sigma_w) \subseteq dom(\sigma_u)$ holds. For a tree $t = (\sigma, l_t)$, $u(t)$ denotes the tree $(\sigma_u @ l_t, l_t)$ and $dom(\sigma) \subseteq dom(\sigma_u @ l_t)$ may not hold anymore[4]. Given two stores $\sigma$ and $\sigma'$, two locations $l \in \sigma$ and $l' \in \sigma'$ are said to be *value equivalent*, written $(\sigma, l) \cong (\sigma', l')$, iff the two trees $\sigma @ l$ and $\sigma' @ l'$ are isomorphic (they possibly differ only in terms of locations). We write $(\sigma, L) \cong (\sigma', L')$ to indicate value equivalence on location sequences $L = (l_1, \ldots, l_n)$ and $L' = (l'_1, \ldots, l'_n)$, with $l_i \in \sigma$ and $l'_i \in \sigma'$, and holding iff $(\sigma, l_i) \cong (\sigma', l'_i)$ for $i = 1..n$.

**DEFINITION 2.4** (INDEPENDENCE). *Let $\sigma$ be a store and $\gamma$ a variable environment. A query q and an update u are said to be independent wrt $(\sigma, \gamma)$ if*

$$\sigma, \gamma \models q \Rightarrow \sigma_q, L_q  \quad  \sigma, \gamma \models u : \sigma_u  \quad  \sigma_u, \gamma \models q \Rightarrow \sigma'_q, L'_q$$

*implies* $(\sigma_q, L_q) \cong (\sigma'_q, L'_q)$. *Also,* q *and* u *are independent, written* q $\perp\!\!\!\perp$ u*, iff they are independent for any pair* $(\sigma, \gamma)$*. Finally,* q *and* u *are independent wrt the DTD* d*, written* q $\perp\!\!\!\perp_d$ u*, iff for every tree* $t = (\sigma, l_t) \in d$ *and* $\gamma$*, they are independent for* $(\sigma, \gamma)$*.*

As a natural consequence of the fact that XML data are typed by a schema, we assume that our independence analysis is run in a context where all data remain consistent wrt the schema after each update. In case an update entails schema evolution, then a larger task of schema maintenance has to be carried on. This task may imply existing views (queries) to be reformulated in order to be correct wrt the new schema, and thus it is likely to exclude any other kind of schema-based analysis until its completion.

## 3. CHAIN INFERENCE

In this section, we define deduction rules to statically infer chains for query and update expressions. Our system produces chains of different kinds. The classification resembles that of Marian and Simeon in [16] for query path extraction, and is needed due to the fact that different kinds of chains play different roles in the independence analysis.

---

[4] $\sigma_u @ l_t$ discards locations disconnected to $l_t$ after the update.

A query chain belongs to one of the following three disjoint classes:

- *Return chains* type input document nodes (*return nodes*) that are roots of elements returned by the query. All descendants of a return node are in the query result, thus a return chain c implicitly embodies these descendants. Now, if a change of an update $u$ targets a *return node* or some of its ancestors or descendants, query-update independence is not guaranteed.

- *Used chains* type nodes (*used nodes*) belonging to the input document and participating to the query evaluation, without necessarily being part of the result itself. Clearly, if a change of an update $u$ targets a *used node* or some of its ancestors, then query-update independence is not guaranteed.

- *Element chains* type newly constructed elements; an element chain is of the form $a.c'$, where $a$ is the tag of the constructed $a$ element. Extracting these chains is important for the precision of the independence analysis (see example below).

For updates, we have one class of chains:

- The purpose of an *Update chain*, denoted by $c{:}c'$, is twofold: c types nodes $l$ whose content may be changed by the update and $c'$ types descendants of $l$ (either introduced or removed by the update) involved in the changes. For example, given $c{:}c'$, independence is not guaranteed if a query returns an element whose root is typed by $c.c''$, with $c''$ a prefix of $c'$.

Let us now illustrate why element chains are necessary for a precise independence analysis. Consider the following update over the well-known XQuery Use Cases DTD [1]:

```
for x in //book return
insert <author>q'</author> into x
```

Here the source expression is an element query, for which we infer element chains of the form $author.c'$, with $c'$ a chain inferred for $q'$. The update chain $bib.book{:}author.c'$ is obtained by concatenation of the chain $bib.book$ associated with the target expression x, and the chain $author.c'$. This allows one to conclude independence wrt the query $//title$, whose unique return chain is $bib.book.title$ (forasmuch as $title$ element is never a descendant of an $author$ element): the update chain is not a prefix of the query chain and vice-versa.

Now, let us do the analysis without considering element chains: for the source expression $<author>q'</author>$, the best that can be done is to infer the chain $author.c:$, telling that something happens beneath book elements. As a consequence, we would not deduce the independence.

In the presence of nested element construction, the same remark holds. In the previous example, if $q'$ is

$$<first>Umberto</first>, <second>Eco</second>$$

then by composing element chains during the inference, we end up with the following two update chains $bib.book{:}author.first.S$ and $bib.book{:}author.second.S$. Indeed, this is necessary to exclude independence wrt the query $//author/email$ (assuming the DTD allows for email elements into author elements).

### 3.1  Chain Inference for XPath Steps

The definition of our chain inference system makes the assumption that the inference is made starting from an input set of chains C. This set can be either $C_d$ (infinite analysis) or a finite subset of $C_d$ (finite analysis). We would like to stress that assuming a pre-computed chain set is only made to ease the formal presentation. Any reasonable implementation can avoid this, by inferring chains on the fly (see Section 6).

The first ingredient for query/update chain inference is chain inference for a single XPath step. We first define chain inference for axes, and then for node tests.

Axis chain inference aims at inferring all chains that can be generated by axis navigation, in a d instance, starting from a node typed by a chain $c \in C$. Chain inference rules strictly mimic XPath semantics of axes, and are defined below (notice that $c'$ can be empty):

$$\mathbf{A}_C(c, \mathtt{self}) \stackrel{def}{=} \{\, c\,\}$$
$$\mathbf{A}_C(c, \mathtt{child}) \stackrel{def}{=} \{\, c.\alpha \mid c.\alpha \in C\,\}$$
$$\mathbf{A}_C(c, \mathtt{descendant}) \stackrel{def}{=} \{\, c.c' \mid c.c' \in C,\ c' {\neq} \epsilon\,\}$$
$$\mathbf{A}_C(c, \mathtt{descendant-or-self}) \stackrel{def}{=} \{\, c.c' \mid c.c' \in C\,\}$$
$$\mathbf{A}_C(c, \mathtt{parent}) \stackrel{def}{=} \{\, c' \mid c = c'.\alpha\,\}$$
$$\mathbf{A}_C(c, \mathtt{ancestor}) \stackrel{def}{=} \{\, c' \mid c = c'.c'',\ c'' {\neq} \epsilon\,\}$$
$$\mathbf{A}_C(c, \mathtt{ancestor-or-self}) \stackrel{def}{=} \{\, c' \mid c = c'.c''\,\}$$

In the following rules, with a little abuse of notation, given a chain c on d, we use $\mathbf{d}(c)$ to indicate either the regular expression $d(a)$, if $c = c'.a$, or the empty regular expression $\epsilon$, if $c = c'.\mathtt{S}$. Chain inference for preceding/following-sibling axes is defined as follows.

$$\mathbf{A}_C(c, \mathtt{following-sibling}) \stackrel{def}{=} \{c_1.\beta \in C \mid c {=} c_1.\alpha,\ \alpha <_{d(c_1)} \beta\}$$
$$\mathbf{A}_C(c, \mathtt{preceding-sibling}) \stackrel{def}{=} \{c_1.\alpha \in C \mid c {=} c_1.\beta,\ \alpha <_{d(c_1)} \beta\}$$

The relation $<_r$ is such that for all $\alpha, \beta \in \Sigma \cup \{\mathtt{S}\}$, $\alpha <_r \beta$ holds if there exists a word $u$ belonging to the language generated by $r$ in which an $\alpha$ occurs before a $\beta$. This relation can be easily defined by structural induction on $r$ (see [9]). For instance, we have $<_{a,(b\mid c)*} = \{(a,b),\ (a,c),\ (b,c),\ (c,b),\ (c,c),\ (b,b)\}$.

Rules for node-test chain inference are straightforward:

$$\mathbf{T}_C(c, \mathtt{node()}) \stackrel{def}{=} \{\, c\,\}$$
$$\mathbf{T}_C(c.\alpha, a) \stackrel{def}{=} \{\, c.\alpha \mid \alpha = a\,\}$$
$$\mathbf{T}_C(c.\alpha, \mathtt{text()}) \stackrel{def}{=} \{\, c.\alpha \mid \alpha = \mathtt{S}\,\}$$

**LEMMA 3.1** (SOUNDNESS OF STEP CHAINS). *Let $t \in \mathtt{d}$ be a tree and $l_x \in dom(t)$. If $\sigma, (\mathtt{x} := l_x) \models \mathtt{x}/\mathtt{axis}{::}\phi \Rightarrow \sigma, L$ then[5] for each $l \in L$ we have:* $\mathtt{c}_l^\sigma \in \mathbf{T}_C(\mathbf{A}_C(\mathtt{c}_{l_x}^\sigma, \mathtt{axis}), \phi)$.

The proof of soundness is reported in [9]. Step chain inference is also minimal for any d, see [9] for further details.

### 3.2  Chain Inference for Queries

Inference rules for queries are presented in Table 1. As usual, a variable environment $\Gamma$ associates each query free-variable x with a set $\Gamma(\mathtt{x})$ of chains, typing nodes that can be assigned to the variable during query evaluation.

Query rules prove judgements of the form:

$$\Gamma \vdash_C \mathtt{q} : (\mathtt{r}; \mathtt{v}; \mathtt{e})$$

meaning that starting from $\Gamma$ and C, the chain inference produces the sets r, v and e, respectively containing the return, used and element chains for q.

---

[5] Notice here that step evaluation does not change $\sigma$.

**Table 1: Chain Inference Rules for Queries**

In the rules, $\bar{\tau}$ denotes all descendant chains of chains in the set $\tau$ wrt $C$:

$$\bar{\tau} \overset{def}{=} \{ c.c' \mid c \in \tau, \ c.c' \in C \}$$

All the rules mimic query semantics [12, 4]. We only comment on the main ones. Rules (FOR) and (LET) are very similar, thus we only comment on the (FOR) rule. It performs an iteration on the set of *return* chains inferred for $q_1$. Return chains for $q_1$ are then converted into used chains. This is needed because chain inference is a bottom-up process: inside $q_1$ a path expression is seen as a query producing a *result* (and as such it *locally* produces return chains), while it only selects nodes to be *used* in the outer iteration `for x in` $q_1$ `return` $q_2$.

In the (FOR) rule, irrelevant chains are filtered out. To illustrate, consider the query

```
for x in //node() return if x/b then x/a
```

The chain inference, thanks to chain filtering, only produces used chains that lead to either an $a$ or a $b$ node. Otherwise, the set of *all* possible chains generated by the subquery `//node()` would be inferred as used chains, for the whole query; as a dramatic consequence, the query would be considered as dependent wrt almost every update.

The rule (STEPF) produces return chains, those pointing to nodes returned by the forward XPath step. The rule (STEPUH) is similar, and deals with upward and horizontal axes. It also produces used chains, by filtering only those bound to the step variable and leading to new result chains according to the step navigation. This is needed since return chains produced by an horizontal/upward step may not contain as a prefix the used chain in $\Gamma(x)$ from which they have been generated. E.g., for the DTD $d = \{a \leftarrow (b+, c*)\}$, and the query `/a/b//following-sibling::c`, we infer $a.b$ as a used chain, and $a.c$ as a return chain.

Element queries `<a>q</a>` are dealt with by rule (ELT). This rule infers element chains of the form $a.c$, where $c$ is obtained from either an element or return chain of $q$. The rule also infers used chains by collecting: *i*) used chains of $q$, and *ii*) used chains *obtained* from return chains of $q$. To this end $\bar{r}'$ is used to extend returned chains of the inner query. This return-to-used chain conversion is needed to correctly handle nested element construction. For instance, consider the following query $q = <r1>q'</r1>$, where

$$q' = (\texttt{x}/a \ , \ <r2>\texttt{x}/b</r2>)$$

Element chains for $q$ are inferred in terms of chains for $q'$. So, element chains for $q$ are $r1.a$ and $r1.r2.b$, assuming that for $q'$ the return chains are $c.a$ (for $\texttt{x}/a$) and $r2.b$ (for $<r2>\texttt{x}/b</r2>$). In order to avoid ending up with a wrong element chain $a.b$ for $q$, the return chain $c.b$ for $\texttt{x}/b$ also does not have to be considered as a return chain for $q'$ as well. This is handled by the return-to-used conversion of the return chain $c.b$ when inferring chains for $<r2>\texttt{x}/b</r2>$ (and hence for $q'$). It is worth stressing that if we just convert return chains to used ones without the extension $\bar{r}$, then we lose their semantic property of representing entire subtrees of data. Notice that this extension is needed for the purpose of the formal presentation although any efficient implementation can avoid performing these extensions by using intensional representations.

The rule (TEXT) deals with expressions building new text nodes. The rule infers $S$ as an element chain[6].

## 3.3  Chain Inference for Updates

As seen before, update chains are of the form $c:c'$. Essentially, the *prefix* $c$ types updated nodes, that are nodes whose children are modified by the update, while the *suffix* $c'$ types modified children or new descendants. Update chains are inferred by rules in Table 2 (only main rules are reported; see [9] for the full set of rules). Chain inference for insert-into expressions (position ranges over into, first and last) is specified by the rule (INSERT-1). For any chain $c:c'$ inferred, the prefix $c$ is a return chain of the target query $q_0$ (typing the insertion point), while the suffix $c'$ is either a return or element chain of the source expression $q$ typing a branch of a node element returned by $q$ itself; this element can either be a new one or a sub-element of the input document; in both cases the suffix chain

---

[6]For simplicity, we preferred not to use a 5th class of chains.

$$\frac{\Gamma \vdash_\mathsf{c} \mathsf{q} : (\mathsf{r}; \mathsf{v}; \mathsf{e}) \qquad \Gamma \vdash_\mathsf{c} \mathsf{q}_0 : (\mathsf{r}_0; \mathsf{v}_0; \mathsf{e}_0) \qquad pos \in \{\mathtt{into}, \mathtt{first}, \mathtt{last}\}}{\begin{array}{c} U = \{\ \mathsf{c:c'} \mid \mathsf{c} \in \mathsf{r}_0,\ \mathsf{c'} \in \mathsf{e}\ \} \ \cup \ \{\ \mathsf{c}{:}\alpha.\mathsf{c''} \mid \mathsf{c} \in \mathsf{r}_0,\ \mathsf{c'}.\alpha \in \mathsf{r},\ \mathsf{c'}.\alpha.\mathsf{c''} \in C\ \} \\ \hline \Gamma \vdash_\mathsf{c} \ \mathtt{insert}\ \mathsf{q}\ pos\ \mathsf{q}_0 : U \end{array}} \text{(INSERT-1)}$$

$$\frac{\Gamma \vdash_\mathsf{c} \mathsf{q} : (\mathsf{r}; \mathsf{v}; \mathsf{e}) \qquad \Gamma \vdash_\mathsf{c} \mathsf{q}_0 : (\mathsf{r}_0; \mathsf{v}_0; \mathsf{e}_0) \qquad pos \in \{\mathtt{after}, \mathtt{before}\}}{\begin{array}{c} U = \{\ \mathsf{c:c'} \mid \mathsf{c}.\alpha \in \mathsf{r}_0,\ \mathsf{c'} \in \mathsf{e}\ \} \ \cup \ \{\ \mathsf{c}{:}\beta.\mathsf{c''} \mid \mathsf{c}.\alpha \in \mathsf{r}_0,\ \mathsf{c'}.\beta \in \mathsf{r},\ \mathsf{c'}.\beta.\mathsf{c''} \in C\ \} \\ \hline \Gamma \vdash_\mathsf{c} \ \mathtt{insert}\ \mathsf{q}\ pos\ \mathsf{q}_0 : U \end{array}} \text{(INSERT-2)}$$

$$\frac{\Gamma \vdash_\mathsf{c} \mathsf{q}_0 : (\mathsf{r}_0; \mathsf{v}_0; \mathsf{e}_0)}{\begin{array}{c} U = \{\ \mathsf{c}{:}\alpha \mid \mathsf{c}.\alpha \in \mathsf{r}_0\ \} \\ \hline \Gamma \vdash_\mathsf{c} \ \mathtt{delete}\ \mathsf{q}_0 : U \end{array}} \text{(DELETE)} \qquad \frac{\Gamma \vdash_\mathsf{c} \mathsf{q}_0 : (\mathsf{r}_0; \mathsf{v}_0; \mathsf{e}_0)}{\begin{array}{c} U = \{\ \mathsf{c}{:}\alpha \mid \mathsf{c}.\alpha \in \mathsf{r}_0\ \} \ \cup \ \{\ \mathsf{c}{:}b \mid \mathsf{c}.\alpha \in \mathsf{r}_0\ \} \\ \hline \Gamma \vdash_\mathsf{c} \ \mathtt{rename}\ \mathsf{q}_0\ \mathtt{as}\ b : U \end{array}} \text{(RENAME)}$$

$$\frac{\Gamma \vdash_\mathsf{c} \mathsf{q} : (\mathsf{r}; \mathsf{v}; \mathsf{e}) \qquad \Gamma \vdash_\mathsf{c} \mathsf{q}_0 : (\mathsf{r}_0; \mathsf{v}_0; \mathsf{e}_0)}{\begin{array}{c} U = \{\ \mathsf{c}{:}\alpha \mid\ \mathsf{c}.\alpha \in \mathsf{r}_0\ \} \ \cup \ \{\ \mathsf{c}{:}\beta.\mathsf{c''} \mid \mathsf{c}.\alpha \in \mathsf{r}_0,\ \mathsf{c'}.\beta \in \mathsf{r},\ \mathsf{c'}.\beta.\mathsf{c''} \in C\ \} \ \cup \ \{\ \mathsf{c:c'} \mid\ \mathsf{c} \in \mathsf{r}_0,\ \mathsf{c'} \in \mathsf{e}\ \} \\ \hline \Gamma \vdash_\mathsf{c} \ \mathtt{replace}\ \mathsf{q}_0\ \mathtt{with}\ \mathsf{q} : U \end{array}} \text{(REPLACE)}$$

**Table 2: Chain Inference Rules for Updates**

corresponds to inserted data. Rule (INSERT-2) is similar, and deals with insert-before/after updates. Inference for delete expressions is defined by the (DELETE) rule, which simply puts the separator ':' just before the last symbol of a return chain of the target query. A delete chain $\mathsf{c}{:}\alpha$ captures that a node typed by $\mathsf{c}$ has a child labeled by $\alpha$ which may be deleted. Similarly, the (RENAME) rule infers chains $\mathsf{c}{:}\alpha$ where $\alpha$ is the type of the target node before renaming, but it also produces chains $\mathsf{c}{:}b$ typing renamed nodes. The rule for replace expressions (REPLACE) is built on the same principles as (INSERT-1) and (DELETE) rules.

### 3.4 Soundness of Chain Inference

From now on, we consider given a DTD $\mathsf{d}$, and some valid document $t{=}(\sigma, l_t){\in}\mathsf{d}$. For a query $\mathsf{q}$, we assume:

$$\sigma, \gamma \ \models\ \mathsf{q} \ \Rightarrow\ \sigma_\mathsf{q}, L_\mathsf{q} \qquad \text{and} \qquad \Gamma \vdash_{\mathsf{c}_d} \mathsf{q} : (\mathsf{r}; \mathsf{v}; \mathsf{e})$$

For an update $\mathsf{u}$, we assume:

$$\sigma, \gamma \ \models\ \mathsf{u} \ \Rightarrow\ \sigma_w, w \quad \sigma_w \vdash w \rightsquigarrow \sigma_\mathsf{u} \quad \text{and} \quad \Gamma \vdash_{\mathsf{c}_d} \mathsf{u} : U.$$

Recall that $\mathsf{u}(t)$ denotes the tree $(\sigma_\mathsf{u}@l_t, l_t)$. Also, for the sake of simplicity, queries and updates are assumed to be quasi-closed: they contain only one free variable $\mathsf{x}$ initially bound to the root of the input XML tree (see [9] for the general case). It means that $\gamma{=}\{\mathsf{x} \mapsto l_t\}$ for query and update evaluation, and $\Gamma{=}\{\mathsf{x} \mapsto \mathsf{d}_s\}$ for static chain inference.

*Soundness of query chain inference.* Proving soundness of query chain rules consists of proving that, for any schema instance, any node used or built by the query $\mathsf{q}$ is captured (typed) by the chains inferred for $\mathsf{q}$. The proof relies on the notion of XML projection [16, 7].

A tree $t'$ is a projection of $t$, denoted $t' \preceq t$, if $t'$ is obtained from $t$ by discarding some subtrees. A projection of a tree $t$ can be obtained from a set $\mathcal{L}{\subseteq}dom(\sigma)$, where $\mathcal{L}$ is non-empty and upward closed with respect to the $\sigma$ parent-child relationship[7]. For a sequence of locations $L$, we define $L_{|\mathcal{L}}$ as the subsequence of $L$ containing only $\mathcal{L}$ identifiers, and preserving $L$ ordering. Then a projection of $t$ wrt a set $\mathcal{L}$ is defined as $t_{|\mathcal{L}}{=}(\sigma_{|\mathcal{L}}, l_t)$ where

$$\sigma_{|\mathcal{L}} \ \overset{def}{=} \ \{\ l \leftarrow a[L_{|\mathcal{L}}] \mid l{\in}\mathcal{L},\ (l \leftarrow a[L]){\in}\sigma\ \} \ \cup \ \{\ l \leftarrow \mathsf{s} \mid l{\in}\mathcal{L},\ (l{\leftarrow}\mathsf{s}){\in}\sigma\ \}$$

We say that $t_{|\mathcal{L}}$ is a $\mathsf{q}$-projection of $t$ if, assuming that $\sigma_{|\mathcal{L}}, \gamma \models \mathsf{q} \Rightarrow \sigma', L'$ we can conclude $(\sigma_\mathsf{q}, L_\mathsf{q}) \cong (\sigma', L')$. Given a set of

[7] $\forall l\ (l{\in}\mathcal{L} \wedge (l'{\leftarrow}a[L]){\in}\sigma \wedge l{\in}L) \ \Rightarrow\ l'{\in}\mathcal{L}.$

chains $\tau$, the set $\mathcal{L}_\tau^t$ of locations in $t{=}(\sigma, l_t)$ typed by chains in $\tau$ is defined as:

$$\mathcal{L}_\tau^t \ \overset{def}{=} \ \{\ l \mid l{\in}dom(\sigma),\ \mathsf{c}_l^\sigma.\mathsf{c} \in \tau\ \}$$

Finally, $t_{|\mathcal{L}}$ is a *minimal* $\mathsf{q}$-projection of $t$ if none of the strict projections of $t_{|\mathcal{L}}$ is a $\mathsf{q}$-projection. Note that, $t'$ is a $\mathsf{q}$-projection of $t$ provided that $t_{|\mathcal{L}} \preceq t'$, for $t_{|\mathcal{L}}$ a minimal $\mathsf{q}$-projection. A minimal projection is not unique, due to the query language considered.

The following theorem formally states that chains inferred for a query $\mathsf{q}$ cover the structure of data relevant for the query, and newly constructed elements.

**THEOREM 3.2. (SOUNDNESS OF QUERY CHAINS)**

1. *If $t'$ is a minimal $\mathsf{q}$-projection of $t$, then $t' \preceq t_{|\mathcal{L}_{\mathsf{r}\cup\mathsf{v}}^t}$*

2. *If $t'$ is the subtree of $\sigma_\mathsf{q}$ rooted at $l'{\in}L_\mathsf{q}\backslash dom(\sigma)$ then $t' \preceq t'_{|\mathcal{L}_{\mathsf{e}}^{t'}}$*

The first item of Theorem 3.2 states that chain inference is sound for used and return chains: a projection of any valid input tree made in terms of used and return chains includes every minimal $\mathsf{q}$-projection, hence preserves query semantics (the projection contains all the query needs for its evaluation). The second item is dedicated to element chain inference which is one of the key feature of our query-update analysis as already illustrated. Intuitively, this statement says that if element chains are used to project newly constructed elements (notice that $l'{\in}L_\mathsf{q}\backslash dom(\sigma)$) no node is pruned out, so element chains cover all possible chains in new elements of the query result.

*Soundness of update chain inference.* Proving update chain soundness consists in establishing a link between $i)$ nodes in the stores $t$ and $\mathsf{u}(t)$ that are *involved* in the changes (deletion, insertion, renaming and replacements) made by $\mathsf{u}$ and $ii)$ nodes in these trees which are captured (typed) by the chains statically inferred for $\mathsf{u}$.

**DEFINITION 3.3 (INVOLVED LOCATION).** *We say that the update $\mathsf{u}$ involves the location $l{\in}dom(\sigma_w)$ if $l$ is either the target location of an elementary delete, rename or replace command in $w$, or a critical location or a descendant of a critical location, where a critical location is a location in the source list $L$ of a command $\mathtt{ins}(L, \_, \_)$ or $\mathtt{repl}(\_, L)$ in $w$.*

Note that an involved location may belong to the initial tree $t$ but not to the updated tree $\mathsf{u}(t)$ and conversely. It may also, of course, belong to both trees. The theorem below states that all locations involved by the update $u$ are typed by chains inferred from $u$.

**Theorem 3.4** (Soundness of Update Chains). *If $l$ is a location in $t$, i.e. $l \in dom(\sigma)$ (respectively a location in $u(t)$, i.e. $l \in dom(\sigma_u @ l_t)$) and the update $u$ involves $l$ then there exists $c:c' \in U$ such that $c_l^\sigma = c:c'$ (respectively $c_l^{\sigma_u} = c:c'$) where $c' \neq \epsilon$.*

In the above statement, in case a location $l$ belongs both to $t$ and $u(t)$, it may be that the chain typing $l$ in $t$ is different from the chain typing $l$ in $u(t)$ (e.g., due to renaming).

Although we made the assumption that the update expression is preserving the schema, it is worth noticing that Theorem 3.4 holds also for updates violating schema constraints ($u(t) \notin d$), since chains corresponding to deleted or inserted nodes are always traced by the system regardless of correctness wrt the schema.

## 4. INFINITE ANALYSIS

The notion of query-update independence $q \perp\!\!\!\perp_d u$ (Definition 2.4) is based on the semantics of $q$ and $u$, and involves all possible $d$ instances. The static counterpart of this notion is now proposed and is of course based on query and update chain inference. As chain inference depends on a set $C$ of chains, we first introduce a general static notion of $C$-independence.

Given two sets of chains $\tau_1$ and $\tau_2$, the set of conflicting pairs of chains for $\tau_1$ and $\tau_2$ is defined by:

$$\text{confl}(\tau_1, \tau_2) \overset{def}{=} \{ (c_1, c_2) \mid c_1 \in \tau_1,\ c_2 \in \tau_2,\ c_1 \preceq c_2 \}$$

**Definition 4.1** (C-independence). *A query $q$ and an update $u$ are $C$-independent, denoted by $q \perp_C u$, if provided that $\Gamma \vdash_C q : (r;\ v;\ e)$ and $\Gamma \vdash_C u : U$, we have:*

$$\text{confl}(r, U) = \text{confl}(U, r) = \emptyset.$$

The main result of this section states that, when $C$ is taken as the set $C_d$ of chains generated for the DTD $d$, $C$-independence implies $q \perp\!\!\!\perp_d u$ independence.

**Theorem 4.2** (Soundness of $C_d$ independence).

$$q \perp_{C_d} u \quad implies \quad q \perp\!\!\!\perp_d u$$

In order to prove Theorem 4.2, the following property is used; it is a consequence of soundness of chain inference (Theorems 3.2 and 3.4). Next, $\mathcal{I}_U^t$ denotes the set of nodes in a tree $t$ typed by update chains in $U$:

$$\mathcal{I}_U^t \overset{def}{=} \{ l \in dom(t) \mid c_l^\sigma = c:c' \in U,\ c' \neq \epsilon \}$$

**Proposition 4.3.** *If $q \perp_{C_d} u$, then we have:*

$$\mathcal{I}_U^t \cap \mathcal{L}_{R \cup v}^t = \mathcal{I}_U^{u(t)} \cap \mathcal{L}_{R \cup v}^{u(t)} = \emptyset$$

This proposition states that $C_d$-independence implies that nodes typed by query chains are disjoint from nodes typed by update chains. The proof is reported in [9].

As already stated, updates are assumed to preserve the schema. The above theorem needs this assumption in order to correctly use *query* chains in the independence analysis. Actually, if deletions violate the schema (a mandatory node is deleted), the $\perp_{C_d}$ is still sound. The problem comes from insertions creating *new* chains (not belonging to $C_d$) because they are not considered during chain inference for queries. As a consequence, the analysis made to check $\perp_{C_d}$ could miss conflicting chains. Extending our technique so as to capture schema evolution is left as future work.

## 5. FINITE ANALYSIS

The notion of $C_d$-independence (Definition 4.1) cannot be directly used to define a terminating decision algorithm, because for DTDs with vertical recursion the sets of inferred chains can be infinite. In this section we show how to finitely approximate sets of inferred chains so that $C_d$-independence can be detected in finite time.

One feature of chains generated by a recursive DTD $d$ is that some of them contain multiple occurrences of (recursively defined) tags. So one way to characterize a finite set of $d$-chains is to restrict to chains having at most $k$ occurrences of each tag. Hereafter, these chains are called $k$-chains, and for any set of chains $\tau$, its subset of $k$-chains is denoted by $\tau^k$. Thus, $C_d^k$ denotes the set of $k$-chains generated by $d$.

As illustrated next, a *multiplicity value* $k$ can be inferred from the query $q$ and update $u$, so that independence according to inferred chains in $C_d$ is equivalent to independence according to inferred chains in $C_d^k$. The value $k$ is derived by a two-steps static analysis.

Given an expression $\text{exp}$, being either a query $q$ or an update $u$, the first step associates a value $k_{\text{exp}}$ to $\text{exp}$ such that the set of $k_{\text{exp}}$-chains inferred for $\text{exp}$ is *representative* of all possible inferred chains for $\text{exp}$. Intuitively, the representative set of inferred chains for an expression synthesizes all possible inferred chains: any possible inferred chain can be mapped to a chain in the representative set by some folding transformations, according to recursive definitions in the DTD. The second step infers a value $k$ from the values $k_q$ and $k_u$, such that the search of conflicting chains decisive for statically detecting $q$-$u$ independence can be safely done in the finite set of inferred $k$-chains.

Inferring the values $k_q$ and $k_u$ mainly depends on navigational properties of the XPath expressions occurring in the query and update. Thus, we start the discussion by focusing on XPath expressions, and then consider FLWR expressions.

*Dealing with child, self and parent.* In this case, a good choice for $k_p$ is the maximal tag frequency in the path $p$. Consider the following recursive DTD $d_1$:

$$r \leftarrow a \qquad b, c, e \leftarrow f \qquad a \leftarrow (b, c, e)* \qquad f \leftarrow a, g$$

For the path $p = /r/a/b/f/a$ the maximal tag frequency is 2, and indeed 2-chains include the representative chain $r.a.b.f.a$ (the only chain inferred for this path); the same holds for the navigational path $/r/a/b/f/a/\texttt{parent}::f$ (note here that the 2-chain $r.a.b.f.a$ is a used chain). Similarly, for the path $/r/a/b/f/*$ we choose $k_p = 2$, since the wildcard $*$ stands for any label.

*Dealing with descendant and ancestor.* When a path $p$ makes use of the $\texttt{descendant}$ axis, the length of inferred chains are totally unrelated to the length of $p$ (e.g., consider $/\texttt{descendant}::b$ over $d_1$). This is what led us to reason in terms of tag frequency rather than path length. Furthermore, such a path can lead us to infer an infinite number of chains over a recursive DTD. To generate a finite set of representative chains, the value $k_p$ is determined by taking into account the number of descendant axes occurrences in $p$.

To illustrate, we still consider the schema $d_1$, and observe that the type $a$ is defined in terms of $b$, $c$ and $e$, and vice versa. In a valid document instance, a $b$ node can be a descendant of a $c$ node, and vice versa, along the same chain of the tree. In addition, a chain connecting $b$ and $c$ nodes always contains an intermediate $a$ label, which also occurs before the first occurrence of a $b$, $c$ or $e$ label. As a consequence, for the following path $p$

$$/\texttt{descendant}::b/\texttt{descendant}::c/\texttt{descendant}::e$$

Over the DTD $d_1$, the shortest chain that is inferred for the path p is $r.a.b.f.a.c.f.a.e$, a 3-chain. Simple tag frequency, like for the previous cases, would lead to $k_p=1$. This is not satisfactory because no chain is inferred for p starting from 1-chains. To reflect the fact that each recursive axis may permit any tag to repeat once in inferred chains, the correct maximal tag frequency we have to consider for the path p is 3; in fact, 3-chains do allow to infer a non-empty set of representative chains. Of course, an XPath expression may combine both recursive and non-recursive axis. In this case, for a path p we obtain $k_p$ as the sum of two components computed independently: the maximal tag frequency for non-recursive steps, and the number of recursive steps in p. As an example, for p=/descendant::b/a/b, we have $k_p=2$ since the maximal tag frequency for the descendant-free part /a/b is 1, and there is 1 descendant step /descendant::b.

Recursive backward axes are handled similarly. Here we have to pay attention to the fact that chains navigated by an `ancestor` step are prefixes of some chains generated by previous steps. Consider p=/descendant::b/ancestor::c. Here $k_p$ has to be such that the *used* chain $r.a.c.f.a.b$ can be generated. Thus, we enforce `ancestor` steps to increment the tag frequency by 1. This is reminiscent of what we have seen before in the case of `descendant`; the way p is processed can be compared to the way the navigational path /descendant::c/descendant::b would be processed, as chains containing $c$ ancestors of $b$ need to be generated for p.

Concerning paths p employing either `descendant−or−self` or `ancestor−or−self`, $k_p$ is computed as for the self-less axes.

*Dealing with sibling axes.* Sibling axes are managed as child and parent axes. Let us consider the recursive schema { $a{\leftarrow}(b, f*)$, $b{\leftarrow}(b|c)*$, $f{\leftarrow}(e, g)$ } and the navigational path /descendant::c/ `following−sibling::b`. For this path, the used 1-chain $a.b.c$ and the return 2-chain $a.b.b$ are the needed chains. The presence of the /descendant::c step entails $k=2$.

*Dealing with FLWR expressions.* Based on concepts previously illustrated, we provide now formal definitions to deal with the general case of FLWR expressions.

As seen before, the computation of $k_{\exp}$ is decomposed into two tasks. The first one determines via the function $\mathcal{F}(a,\exp)$ the frequency of each tag $a{\in}\Sigma$ on the whole expression, in order to derive the maximal frequency. The second task computes via the function $\mathcal{R}(\exp)$ the maximal number of consecutive recursive steps in the whole expression. The value $k_{\exp}$ is the sum of these two values. Formally:

$$k_{\exp} \overset{def}{=} \max\{ \mathcal{F}(a,\exp) \mid a{\in}\Sigma \} + \mathcal{R}(\exp)$$

The functions $\mathcal{F}(a,\exp)$ and $\mathcal{R}(\exp)$ are defined by structural induction in Table 3. When exp is a for/let expression, the value $k_{\exp}$ is specified by summing the sub-expression values. This is motivated by the fact that, for instance, for-expressions are usually used to encode nested iterations performed by XPath paths, like in the query `for x in /a for y in x/b return y`. This leads in some cases to an overestimation of the value $k_{\exp}$ that would be actually sufficient for a finite analysis. For instance, for the query q'

    for x in /a/a return for y in /a/b return x,y

we have $\mathcal{F}(a,q')=3$, while the value 2 would be sufficient. More precision can be obtained by tracing variable bindings in the definition of $\mathcal{F}(,)$. The same argument holds for $\mathcal{R}()$. However, this would make the formalization cumbersome without being a decisive factor for the analysis. Thus, our choice has been guided by simplicity and conciseness of $\mathcal{F}(,)$ and $\mathcal{R}()$ definitions.

| | | |
|---|---|---|
| $\mathcal{F}(a, \exp) \overset{def}{=}$ | 0 | if exp is () or s or exp is x/axis :: $\phi$ and axis recursive or $\phi{\notin}\{a,\texttt{node}()\}$ |
| | 1 | if exp is x/axis :: $\phi$ and axis not recursive and $\phi{\in}\{a,\texttt{node}()\}$ |
| | $\max\{ \mathcal{F}(a, \exp_i) \}$ | if exp is $(\exp_1,\exp_2)$ or (if $\exp_0$ then $\exp_1$ else $\exp_2$) |
| | $\sum_i \mathcal{F}(a, \exp_i)$ | if exp is (for/let x $\exp_1$ return $\exp_2$) or (delete $\exp_1$) or (insert/replace $\exp_1$ $\exp_2$ ) |
| | $\mathcal{F}(a, \exp)$ | if exp is (`<b>`exp`</b>`) or (rename exp as b) and $b{\neq}a$ |
| | $1 + \mathcal{F}(a, \exp)$ | if exp is (`<b>`exp`</b>`) or (rename exp as b) and $b{=}a$ |
| $\mathcal{R}(\exp) \overset{def}{=}$ | 0 | if exp is () or s or x/axis :: $\phi$ and axis not recursive |
| | 1 | if exp is x/axis :: $\phi$ and axis recursive |
| | $\max\{ \mathcal{R}(\exp_i) \}$ | if if exp is $(\exp_1,\exp_2)$ or (if $\exp_0$ then $\exp_1$ else $\exp_2$) |
| | $\sum_i \mathcal{R}(\exp_i)$ | if exp is (for/let x $\exp_1$ return $\exp_2$) or (delete $\exp_1$) or (insert/replace $\exp_1$ $\exp_2$) or (rename $\exp_1$ as b) |

**Table 3:** $\mathcal{F}(,)$ and $\mathcal{R}()$ definition.

The rule for element construction deserves some comment. Note that tags of constructed elements are taken into account. Indeed, these elements can be inserted by an update as children of existing elements, thus generating new chains that can be used by a query. Consider the recursive schema { $a \leftarrow b,\ b \leftarrow b?,\ c?$ } and the following update u:

    for x in /a/b return
        insert  <b><b><c/></b></b> into x

As already outlined, precision of the independence analysis relies (among other things) on the chains generated for element construction. The rules in Table 3 lead to $k_u=3$, and thus the chain $a.b.b.b.c$ is inferred for the finite analysis. Note that tag frequency for rename expression is determined in a similar way: after renaming, the tag frequency may increase, and chains, for the finite analysis, have to capture this change. Other rules are self-explicative.

*Finite independence analysis.* We see now how to use the values $k_q$ and $k_u$ in order to determine a $k$ value such that $C_d$-independence can be detected by restricting to $k$-chains.

Consider q=/descendant::b and u=delete /descendant::c over the previous DTD $d_1$. They are dependent and $k_q=1$, $k_u=1$. We could argue that a sound choice is $k{=}max(k_q,k_u)$, which allows the finite analysis to infer the query chain $r.a.b$ and the update chain $r.a.c$. Unfortunately, these chains do not conflict, and rule out dependence. The problem here comes from the fact that the update may change a descendant of a query returned node, and that $k{=}max(k_q,k_u)$ does not permit to capture this in the finite analysis. To avoid this problem, it is necessary that representative chains that are inferred for the update cover query returned nodes. To this end, while inferring chains for the update u, structural properties of the query q have also to be taken into account. This is obtained by setting $k$ to $k_q + k_u$.

In the remaining part of this section we focus on one of our main results, soundness of the finite analysis.

THEOREM 5.1 (SOUNDNESS OF $C_d^k$ INDEPENDENCE). *Let* d *be a DTD,* q *a query and* u *an update. Let* $k{=}k_q + k_u$ *as defined above. Then:*

$$q \perp_{C_d^k} u \ \ implies \ \ q \perp\!\!\!\perp_d u$$

We focus on soundness because completeness (q $\perp_{C_d}$ u *implies* q $\perp_{C_d^k}$ u) is straightforward, as $C_d^k{\subseteq}C_d$.

We next develop the main steps of the proof of Theorem 5.1. We reason in terms of *dependence*, rather than independence. We prove that $C_d$-dependence implies $C_d^k$-dependence (these notions directly follow from Definition 4.1) by showing that from any pair of chains in $C_d$, witness of dependence, it is possible to identify a pair of $k$-chains in $C_d^k$, witness of dependence.

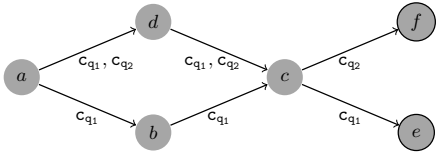

**Figure 2: CDAG for** $q_1, q_2$

The proof is composed of three steps. First, we show that there exists a folding from query chains to $k$-query-chains, for any query $q$ (Lemma 5.2). Then, we show that there exists a folding from query chains to $k$-query-chains also preserving the prefix relation $\preceq$, for any pair of queries $(q, q')$ (Lemma 5.3). Finally, we show that such a folding exists for chains inferred for any query-update pair $(q, u)$ (Theorem 5.1). Proofs are reported in [9].

Given a DTD $d$, we define a folding relation $\hookrightarrow_d \subseteq C_d \times C_d$ as

$$\hookrightarrow_d \overset{def}{=} \quad \{ \ (c_1, c_2) \mid c_1 = c.a.c'.a.c'' \ \land \ c_2 = c.a.c'' \ \}$$

Notice that, above, the symbol $a$ is a recursive type of the schema. We dub $\hookrightarrow_d^*$ the reflexive and transitive closure of $\hookrightarrow_d$.

**LEMMA 5.2** (FOLDING). *For each chain* $c$ *inferred from a query* $q$ *there exists a chain* $c'$ *inferred for* $q$ *such that* $c \hookrightarrow_d^* c'$ *and* $c'$ *is a* $k_q$*-chain.*

When $q$ and $u$ are $C_d$-dependent, at least one of $\text{confl}(U, v)$, $\text{confl}(r, U)$, $\text{confl}(U, r)$ is nonempty (see Definition 4.1). This implies that there exists a conflicting pair of inferred chains, witness of the $C_d$-dependence of $q$ and $u$. As updates are defined *in terms of* queries, the next lemma which focusses on "conflicting" query chains is needed to conclude the proof of Theorem 5.1.

**LEMMA 5.3** (FOLDING AND CONFLICT PRESERVATION). *For each pair of chains* $(c_1, c_2)$ *inferred for queries* $q_1, q_2,$ *and such that* $c_1 \preceq c_2,$ *there exists* $(c_1', c_2')$ *inferred for* $q_1, q_2,$ *such that* $c_1' \preceq c_2'$ *and* $c_i \hookrightarrow_d^* c_i'$ *with* $c_i'$ *a* $(k_{q_1} + k_{q_2})$*-chain* $(i=1, 2)$.

# 6. IMPLEMENTATION AND EXPERIMENTS

## 6.1 Complexity and Implementation

We implemented in Java our technique for independence analysis. The crucial aspect of the implementation concerns the choice of the data structure for representing inferred chains for the query and the update: the overall performance of the analysis depends on this. This is because the number of distinct chains inferred for a single expression can grow exponentially with the size of the expression to analyze[8].

In order to avoid this blow-up we represent a set of inferred chains for the query (or update) as a chain-DAG (CDAG) where common *prefixes* and *suffixes* shared by different chains are merged according to the following principle. A CDAG is rooted at the schema root type, contains no self-loops and meets the following property: for all type $\alpha$ defined in $d$, there is at most one CDAG-node of type $\alpha$ at a distance $h$ from the root. In other words, if two chains happen to have the same type-name in position $h$, they will share a common node in the CDAG at depth $h$. This implies that during chain inference the width of the CDAG is upper-bounded by the schema size.

The CDAG representation requires a small overhead in order to distinguish two chains inferred for distinct sub-expressions, in the case that these chains share some nodes of the graph. To this end, any edge connecting two CDAG-nodes is labeled with a code identifying the query/update expressions that created it during chain inference. These identifiers are necessary to correctly perform chain inference for backward axes, and independence checking as well. For instance, consider the query $q_1, q_2 = //c/e, /a/d/c/f$. Assume $q_1$ and $q_2$ produce $\{a.b.c.e, a.d.c.e\}$ and $\{a.d.c.f\}$. The CDAG representation of these chains is illustrated in Figure 2. First, we clearly see that the merge does not produce non-existing chains as artifacts: by following query codes there is no way to trace a chain $a.b.c.f$. Second, we can observe that if in $q_1, q_2$ we had $q_2 = /a/d/c/f/\text{ancestor}::*$, when inferring chains for the last step of $q_2$, thanks to edge-labeling we avoid to navigate upward parts of the CDAG that have not been generated by $q_2$ (i.e., a $b$ node). Notice that backtracking on unvisited nodes would not affect the correctness of the analysis, but would compromise precision of the independence analysis.

An auxiliary index associates each expression with nodes representing ending points of inferred used and return chains ($e$ and $f$ nodes in Figure 2). Concerning element chains, these are kept in a specific/separate component of the CDAG, in order to distinguish among chains those typing input and those typing constructed data.

The following theorem proves that the chain inference needed for checking $C_d^k$-independence has polynomial complexity.

**THEOREM 6.1.** *By using CDAGs, finite chain inference for a DTD* $d$*, a value* $k$*, and an expression* $\text{exp}$*, can be done in* $O(k^2 \times |d|^4)$ *space and* $O(|\text{exp}| \times k^2 \times |d|^5)$ *time.*

For space reason, the proof is reported in the full version [9]. Here we discuss some cases of practical relevance for which complexity is better than that stated in Theorem 6.1.

When the test condition $\text{node}()$ is not used in XPath steps, then time complexity is $O(|\text{exp}| \times k^2 \times |d|^4)$. This is because each inference step would produce $O(k \times |d|)$ nodes in the CDAG (i.e., at most one for each CDAG level), while with $\text{node}()$ it produces $O(k \times |d|^2)$. Furthermore, if we assume that during chain inference each XPath step can have at most $m$ CDAG nodes as input, time complexity goes down to $O(m \times |\text{exp}| \times k \times |d|^3)$. The value $m$ is likely to be close to 1 for most XPath steps used in practice. This holds in particular for XMark and XPathMark expressions. Another fact observable from such expressions is that they employ a small number of recursive navigations, thus making chain inference doable in $O(|d|^3)$ time.

When $d$ is not recursive, the $k$ value stops being determinant for the analysis since no label repeats twice in any $d$ chain. In this case the number of edges of the CDAG is bound by the size of the parent-child relation induced by the schema[9]. Therefore the spatial complexity goes down to $O(|d|^4)$, while time complexity is $O(|\text{exp}| \times |d|^2)$. If we also assume the absence of the test filtering $\text{node}()$, time complexity is $O(|\text{exp}| \times |d|)$. These restrictions are often met in practice, and in particular by expressions used in our testbed (when the recursive component of the XMark schema is not visited at all by the expression).

---

[8] This happens for schemas that make heavy use of recursive definitions, but also for non-recursive ones, like for instance $d = \{ a_i \leftarrow (b_i, c_i) * \quad b_i, c_i \leftarrow a_{i+1} \ , \ i = 1..n \}$ (for a query $q = //a_n$ the number of inferred chains is $2^n$).

[9] If the parent-child relation has more than $|d|(|d|-1)$ elements the schema is recursive.

Once chain inference is done for a query `q` and an update `u`, independence (Definition 4.1) is checked over the two inferred CDAGs. This check can be done in $O(c \times |q| \times |u|)$ time, where $c$ is the size of the smallest CDAG.

## 6.2 Experiments

We performed extensive experiments by using our Java implementation, in order to measure i) efficiency, ii) precision and iii) scalability of our static analysis. We used two different benchmarks: a first one based on XMark /XPathMark, and a second one, dubbed R-benchmark, we specifically designed to measure iii). Concerning the first one, we used a superset of the *view maintenance benchmark* adopted by Benedikt and Cheney in [6]. Our benchmark is composed of a set of 36 views $v_i$ and a set of 31 updates $u_i$. A view is a query belonging to either the XMark query set $q_1$-$q_{20}$ [20], or to the XPathMark query set $A1$–$A8$/$B1$–$B8$ [13]; $Ai$ queries only use downward axes, whereas $Bi$ queries use upward and horizontal axes as well. Concerning updates, a first set corresponds to those used in [6]; these are derived from the XPathMark query set $A1$–$A8$/$B1$–$B8$ and are of the form $UAi$=`delete Ai` or $UBi$=`delete Bi`. We added a set of 15 updates formed by insert expressions $UI1$–$UI15$, rename expressions $UN1$–$UN5$, and replace expressions $UP1$–$UP5$. These updates have been defined so as to cover all different types of nodes in XMark documents, and in particular those pasted by mutually recursive types. It is worth remarking that, even if not all of the delete-updates of the testbed preserve the schema (see UA4, UA5, UA6, UA7, UA8, UB1, UB5, UB6, UB7, UB8 ), the correctness of our technique is still ensured, since no new chain is created by these expressions. As outlined before, our technique is just unaware of new chains built by breaking schema constraints. In light of this, insert, rename and replace update expressions have been chosen in order to preserve document validity. Before performing the tests, XMark and XPathMark expressions have been opportunely rewritten into expressions belonging to the XQuery fragment we consider (Section 3), as done in [5]. The rewriting essentially consists of: putting predicate conditions in disjunctive form, removing attribute use, and extracting paths from functions calls and arithmetic expressions. Clearly, the rewriting is such that a query and an update are independent if the rewritten query and update are. Due to lack of space, queries and updates are reported in the full version [9].

We used the above described benchmark to measure precision and efficiency of our technique. Concerning the R-benchmark, it is designed for understanding the impact of recursion in the performances of our analysis. It is formed by schemas and expressions with a massive use of recursion; it is described later on.

We ran all tests on a desktop 4-core Intel Xeon 2.13 GHz machine with 8 GB RAM (the JVM was given 2 GB) running Linux. To avoid perturbations coming from system activity, we ran each experiment ten times, discarded the best and the worst performance, and computed the average of the remaining times.

*Runtime on XMark.* We measured the time needed by the static analysis to detect independence of each update wrt the whole set of XMark views. The XMark schema is particularly suitable for testing the performances of our technique since the type dependency graph of this schema contains 5 mutually recursive types that form two cliques of size 2 and 3 respectively. We recall that the execution cost depends on the three parameters $|d|$, $|exp|$ and $k$. In this testbed we have $|d| = 76$, and $|exp| \leq 20$, while multiplicity values $k$ range from 2 to 6. As observed in Section 6.1, in many cases chain inference is made in $O(|exp| \times |d|)$ time.

Time values include the time for CDAGs inference and comparison, for each pair of expressions. Results are collected in Figure 3.a. It shows that the analysis is quite fast: in the worst case the analysis is performed in less than 40 ms for the whole set of views, while the average cost is around 15 ms. According to complexity results of Section 6.1, inference time is influenced by i) the $k$ values needed by a query-update pair and ii) the number of recursive types of the schema effectively unfolded. We see small changes in inference time values according to the $k$ value (e.g., the pair UB1-UB2). Yet, two expressions having the same $k$ value may have different time costs for chain inference, depending on the effective number of recursive types unfolded by the analysis (e.g., the pair UI3-UP3).

Running times obtained from the available OCaml implementation[10] of the analysis presented in [6] are rather close to ours: the average time for analyzing an update vs all of the views is around 10 ms. It is worth observing, that inference time for [6] has no sensible oscillations, while in our case inference time depends on $k$, hence on the query and update expressions. The analysis presented in [6] has time complexity $O((|d|^2 + |q|)^2 + |u|)$, and thus is expected to be faster than our analysis in the presence of recursive schemas. Nevertheless, as shown shortly, our running times remain low enough to ensure high time savings in views maintenance, even when views are defined on relatively small documents.

*Precision on XMark.* Independence (Definition 2.4) is undecidable in general [6], so for the purpose of measuring precision, for each update $u_i$ we manually determined independent pairs $(u_i, v_j)$, details are reported in the full version [9] (note that for most pairs in the considered testbed independence is evident, so this process is much less time consuming than one may guess). We then express precision as the percentage of independent pairs that are deemed independent by our static analysis too. To estimate improvements wrt the alternative schema-based technique [6] we computed the same percentages for that technique by using the public tool[10].

Results are reported in Figure 3.b. Our chain-based analysis turned out to be precise. Percentages go from 72% to 100%, while the average precision is 96%. Also, Figure 3.b shows that the analysis proposed in [6] (that has an average detection of 49%) is always outperformed in terms of precision by our static analysis, and in some cases improvements are huge. This happens in particular for updates UB1, UB5, UB6, UB8 (employing backward and horizontal axes).

For the updates, the over-approximation made by type rules in [6] entails a high number of false negatives. Our chain based inference instead is so precise to avoid most of these false-negatives. In general, improvements in terms of precision go from 8% (UN4) to 96% (UP1), and the average gain is 46%. In particular, precision of our analysis remains high in the presence of views using upward and horizontal axes (XPathMark queries in the group B). These queries are likely to be among the most expensive ones to re-evaluate after document updating.

*Maintenance time on XMark.* We measured time savings obtained by avoiding the re-materialization of views which our analysis deem as independent of an update. We used three XQuery engines: Saxon 9.2EE, BaseX 7.0.1 and QizX 4.4. We considered a 1MB XMark document and we scaled to 10MB and 100MB, in order to measure time savings in real scenarios.

Our test results only take into account query answering time. Full details about engine configurations can be found in [9]. For this experiment only, the JVM was given 4GB of RAM, in order to minimize memory swapping. Results are reported in Figure 3.c.

---

[10] `http://homepages.inf.ed.ac.uk/jcheney/programs`

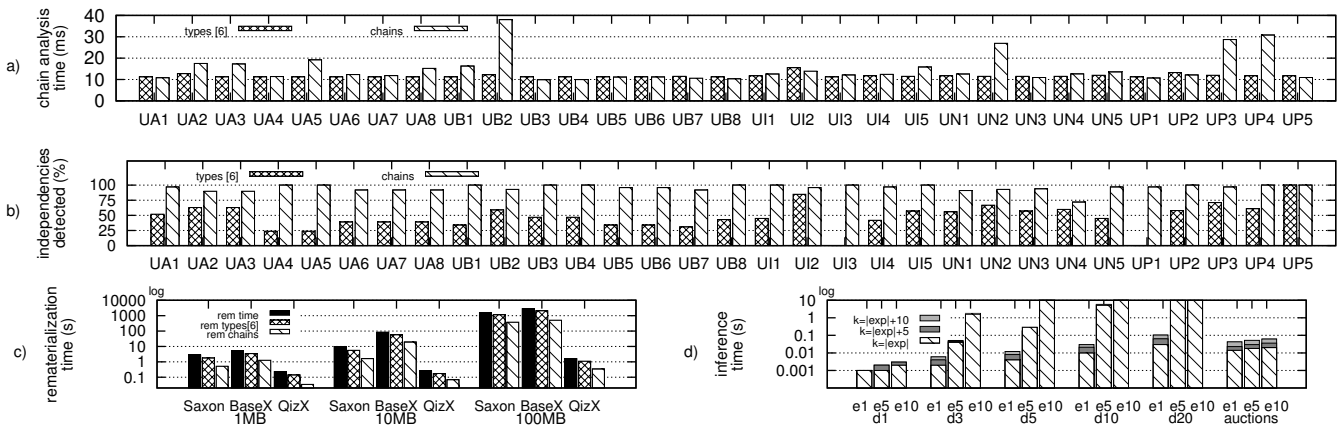

**Figure 3: Test results**

As in [6], for each update $u_i$ we measured the time $r_i$ needed for refreshing *all* the 36 views after the update, and the time $r_i^{type}$ and $r_i^{chain}$ needed to refresh only views that are not deemed as independent by the static analysis of [6] and by ours, respectively. In Figure 3.c, for each of the three used engines we report the averages of all refreshing times $r_i$, $r_i^{type}$ $r_i^{chain}$. As a consequence of time efficiency and precision of our static analysis, even for a relatively small document of 1MB, our independence analysis ensures high time savings for all engines: 82% for Saxon, 75% for BaseX and 85% for QizX. While type based analysis [6] ensures much lower time savings: 36% for Saxon, 31% for BaseX and 37% for QizX.

These percentages are essentially the same as those obtained for 10MB and 100MB documents, both for our technique and for that of [6]. This is because in the considered benchmark, queries that are not statically deemed as independent of an update, and hence refreshed, are the most expensive ones to refresh.

*Scalability on R-benchmark.* The benchmark is composed of a parametric schema $dn$ including $n$ fully-mutually recursive types (each of the $n$ types is defined in terms of all the $n$ types), and a set of XPath expressions $em$, each one consisting of $m$ consecutive `descendant::node()` steps. Parameters $n$ and $m$ allow us to range over several configurations and trace the perimeter of applicability of our technique. We considered four schemas $dn$ with $n$ ranging over $\{1, 3, 5, 10, 20\}$, and, for each schema, three expressions $em$ with $m$ ranging over $\{1, 5, 10\}$. Also, for each expression $em$ we considered $k$ ranging over $\{|em|, |em|+5, |em|+10\}$. Observe that $|dn|=n$ and $|em|=m$.

The schema $d5$ is quite complex, it contains 5 mutually recursive types. We can see from Figure 3.d that even with such complex form of recursion, for $e5$, and for each $k \in \{5, 10, 15\}$, chain inference is still fast (inference time is around a decimal of a second). For schema $d10$, featuring an extremely complex form of recursion, inference time is around five seconds for $e5$, while for $e10$ the time exceeds ten seconds. The same happens for more complex cases.

These test results show that even for forms of recursions that are unlikely to occur in practice (like the $d5$-$e5$ case), chain inference is still fast, while it takes more than one second for extremely complex cases. Figure 3.d also report test results on chain inference of expressions $em$ over the XMark DTD. As it can be seen, if we make a comparison with the $d3$ case (recall that the largest clique has size 3 in the XMark schema) the number of type definitions (76 in this case) have an impact on inference time, since the query expressions make a massive use of `descendant::node()` steps. As already discussed, when such step is not used, inference time drastically reduces, as often happens in practice, and in particular for many XMark/XPathMark expressions.

## 7. EXTENSIONS

*Queries and updates.* While we have considered all update operators made available by XQuery Update Facility, the XQuery fragment we have considered (the same as the one considered in the related approaches [6, 5]) leaves out several query mechanisms. These can be handled by means of two possible methods. The first one is based on query rewriting. A basic form has been used in [5], as well as in our experiments (see Section 6). The second method is based on providing new inference rules. The two methods can be used together, and are both easy to develop, except for user defined recursive functions, whose treatment is beyond the scope of this work since they introduce Turing completeness.

For space reason, details about extensions are given in the full version [9]. Here, we would like to stress that what makes them easy to develop is our static concept of C-independence, based on the notions of used, return and element nodes (Section 3). These are *universal* and *essential* notions, in the sense that, for any kind of query construct that one could think of adding to the framework, analyzing the role of a node with respect to this construct makes the node fall in one of these three categories. Thus, generalizing our framework for a new query construct mainly consists of identifying how used, return and element nodes are determined. This simply requires understanding the standards concerning the query construct semantics, and reusing principles followed in the treatment of the core language in Section 2.

*Schemas.* Concerning schemas, our technique can be extended in order to deal with Extended DTD [14], capturing XML Schema and RelaxNG types.

DEFINITION 7.1. *An Extended DTD is specified by a tuple* $(\Sigma, \Sigma', s, d, \mu)$ *where* $(\Sigma', s, d)$ *is a DTD and* $\mu$ *is a function from* $\Sigma' \cup \{S\}$ *to* $\Sigma \cup \{S\}$ *such that* $\mu(S) = S$.

A tree $t$ is valid wrt $(\Sigma, \Sigma', s, d, \mu)$ if and only if $t' = \mu(t)$ and $t'$ is valid wrt the DTD $(\Sigma', s, d)$. Following [14], in an EDTD we can assume $\Sigma' = \{a_i | a \in \Sigma\}$ and $\mu(a_j) = a$ for all $a_j \in \Sigma'$. This implies that two types differently indexed produce the same label but possibly different content models. That said, it is sufficient to change Definition 2.1, and the definition of $\mathbf{T}_C(c, a)$ with $a \in \Sigma$. Both changes are straightforward. Notice that precision of the inference as well as complexity results remain unchanged for the EDTD case.

Concerning attributes, extensions are straightforward, and actually implemented in our prototype (a simple rule for dealing with the `attribute` axis is needed). Concerning ID/IDREF constraints in DTDs, and key/keyref constraints in XSDs (studied in [2]), we assume they are preserved by updates, as we assume that validity is preserved (Section 2). So, in order to ensure precise and sound independence analysis, chain inference does not need to consider these constraints. Our notion of C-independence only concerns the *type* component of the schema, while these constraints pose restrictions on the values of attributes and elements in a document, and do not impact its structure.

## 8. RELATED WORK

Besides [15] and [6], already discussed, another work quite close to ours is that recently presented by Benedikt and Cheney in [5]. An important contribution of this work was a schema-less framework that factors the problem of independence analysis into two sub-problems: i) statically inferring a set of *destabilizers* queries from a query, and ii) checking whether destabilizers overlap with the target nodes of an update. Precision of the technique mainly depends on the kind of destabilizers that are inferred from XPath steps. To this regard, the inferred destabilizers for steps of the form `x/child::b` include `x/child::*` (a similar inference is made for other downward axes). As a consequence any update touching a non-b node which is a sibling of a $b$ node selected by `x/child::b` would not be detected as independent of `x/child::b`, while it should. In the presence of a schema, our technique detects independence for these cases, thus ensuring a much higher degree of precision. It is worth observing that precision of this destabilizer-based approach could be improved by adopting a different destabilizer inference system, but yet ensuring high precision could be hard since, as shown in [5], there is non elementary algorithm for constructing a minimal static destabilizer.

Type-based projection techniques [7, 3] could be extended to detect query-update independence. However, as type-projectors resemble to types inferred by [6], the extension would not be as precise as our technique. Also, both techniques [7, 3] only consider DTDs, while chain-based analysis works for EDTDs too.

Raghavachari and Shmueli [18] considered a downward subset of XPath, and found fragments for which independence turns out to be either a polynomial or an NP-hard problem; schema information was not considered.

## 9. CONCLUSIONS

We presented a type system able to statically detect XML query-update independence. One of the main feature of the type system is the chain inference component, allowing to infer information at the basis of an highly precise analysis. One of the key contributions of the work is a method to restrict the analysis to a finite set of chains in the presence of recursive schemas. As shown by examples and experiments our technique ensures high improvements in terms of precision wrt the state-of-the art schema-based technique [6].

*Acknowledgments.* We would like to thank George Katsirelos, Asterios Katsifodimos and Carlo Sartiani for helpful discussions and comments on this work. We would also like to thank the VLDB anonymous referees for their useful feedback and suggestions. This work has been partially funded by *Agence Nationale de la Recherche*, decision ANR-08-DEFIS-004.